\documentclass[prl,twocolumn,superscriptaddress,showpacs]{revtex4-1}
\usepackage{amsmath}
\usepackage{amssymb}
\usepackage{xspace}
\usepackage{graphicx}
\usepackage{grffile}
\usepackage{nicefrac}
\usepackage{bbm}

\usepackage{dcolumn}

\usepackage{hyperref}
\hypersetup{colorlinks=true,breaklinks,linkcolor=blue,urlcolor=blue,citecolor=blue}

\usepackage{color}

\def\Im{\hbox{Im}}

\newcommand{\abs}[1]{\ensuremath{\lvert#1\rvert}}

\renewcommand{\Im}{\operatorname{\mathrm{Im}}}

\def\qvec{\mathbf{q}}

\def\kvec{\mathbf{k}}

\def\kv{\kvec}
\def\qv{\qvec}

\begin{document}

\title{From Hubbard bands to spin-polaron excitations in the doped Mott material Na$_x$CoO$_2$}

\author{Aljoscha Wilhelm}
\affiliation{I. Institut f{\"u}r Theoretische Physik, Universit{\"a}t Hamburg,
D-20355 Hamburg, Germany}

\author{Frank Lechermann}
\affiliation{I. Institut f{\"u}r Theoretische Physik, Universit{\"a}t Hamburg,
D-20355 Hamburg, Germany}

\author{Hartmut Hafermann}
\affiliation{Institut de Physique Th\'eorique (IPhT), CEA, CNRS, 91191 Gif-sur-Yvette, France}

\author{Mikhail I. Katsnelson}
\affiliation{Radboud University Nijmegen, Institute for Molecules and Materials,
NL-6525 AJ Nijmegen, The Netherlands}

\author{Alexander I. Lichtenstein}
\affiliation{I. Institut f{\"u}r Theoretische Physik, Universit{\"a}t Hamburg,
D-20355 Hamburg, Germany}

\begin{abstract}
We investigate the excitation spectrum of strongly correlated sodium cobaltate within a realistic many-body description \emph{beyond} dynamical mean-field theory (DMFT). At lower doping around $x$=0.3, rather close to Mott-critical half-filling, the single-particle spectral function of Na$_x$CoO$_2$ displays an upper Hubbard band which is captured within DMFT. Momentum-dependent self-energy effects beyond DMFT become dominant at higher doping. Around a doping level of $x\sim 0.67$, the incoherent excitations give way to finite-energy spin-polaron excitations in close agreement with optics experiments. These excitations are a direct consequence of the formation of bound states between quasiparticles and paramagnons in the proximity to in-plane ferromagnetic ordering.
\end{abstract}

\pacs{
71.27.+a,
72.10.Di,
71.45.-d
}

\maketitle

\begin{figure}[t]
\begin{center}
(a)\hspace*{-0.05cm}\includegraphics*[height=4cm]{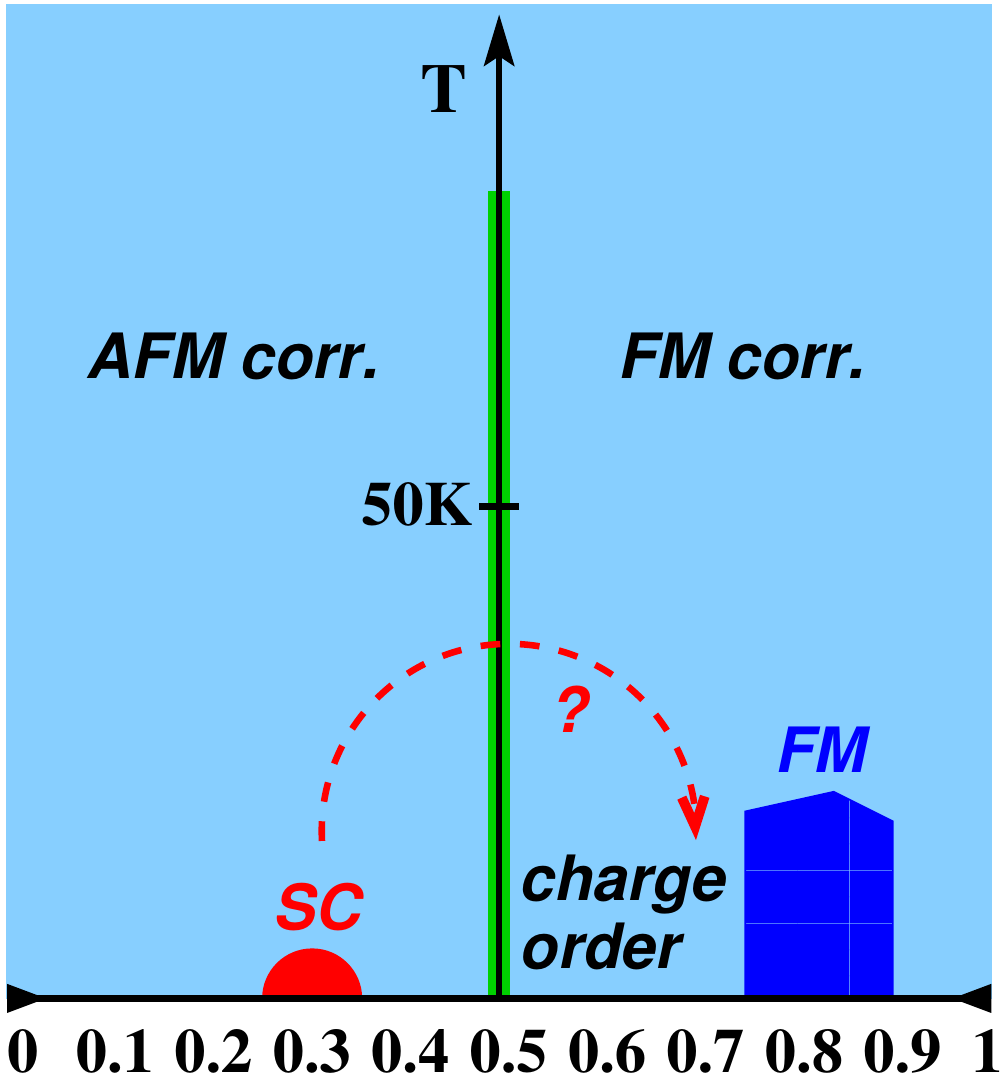}\hspace*{-0.1cm}
(b)\hspace*{-0.25cm}\includegraphics*[height=4cm]{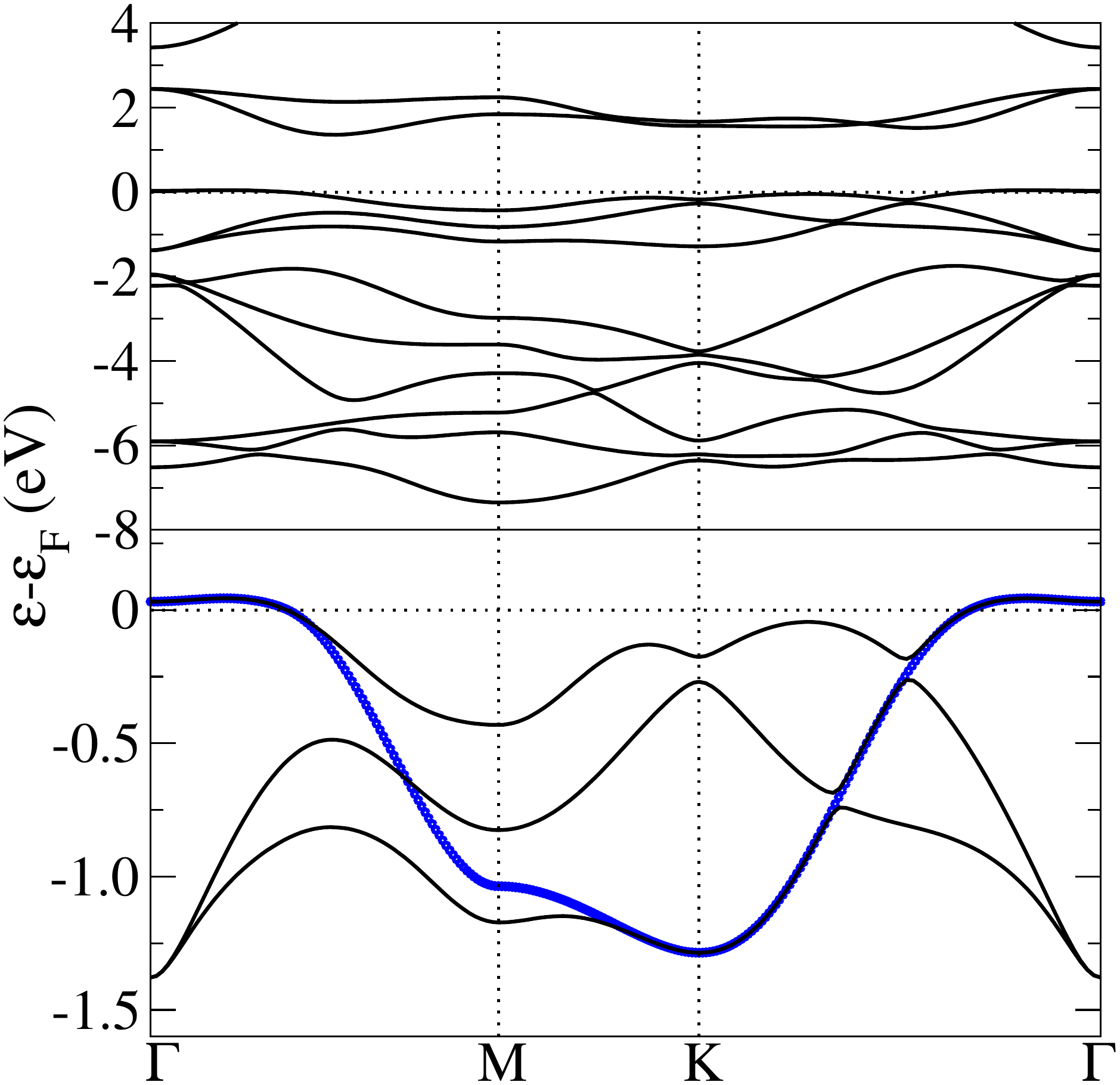}\\[0.2cm]
(c)\includegraphics*[height=3cm]{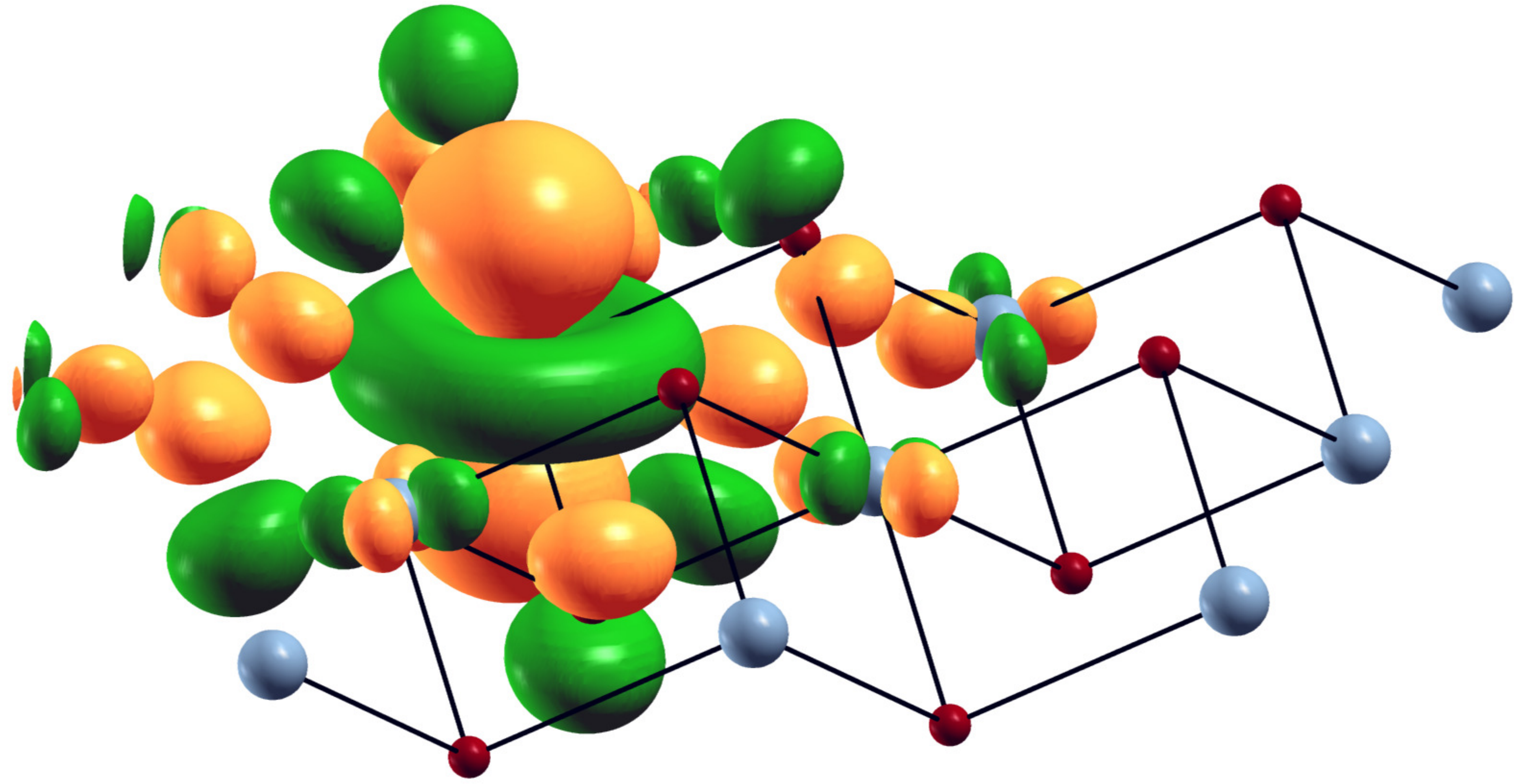}
\end{center}
\caption{(Color online) (a) Schematic $T-x$ phase diagram for the CoO$_2$ planes in
sodium cobaltate. The green bar at $x$=0.5 marks the insulating charge-ordered state and the 
dashed arrow indicates the possibility for superconductivity close to
$x$=0.67~\cite{oht11}. (b) LDA band structure for $x$=0.67. The lower panel shows a blow up of the region around $\varepsilon_{\rm F}$ with the $a_{1g}$-like Wannier band (blue).
(c) Corresponding real-space Wannier orbital in the CoO$_2$ plane (Co: grey, O: red).
\label{material-fig}}
\end{figure}

The relevance of electronic correlation effects beyond the cuprate paradigm has
become obvious in recent years due to issues raised by materials such as cobaltates, iron pnictides/chalcogenides, oxide heterostructures or iridates. The experimental phase diagram of the quasi-two-dimensional sodium cobaltate system Na$_x$CoO$_2$ exhibits nearly all hallmarks of strongly correlated physics (see Fig.~\ref{material-fig}a). Electron doping of stacked triangular CoO$_2$ layers via $x$ Na$^+$ ions results in superconductivity upon intercalation with water~\cite{tak03}, charge disproportionation~\cite{muk05,lan08}, Curie-Weiss behavior~\cite{foo04}, in-plane ferromagnetic (FM) order~\cite{sug03,mot03,boo04} and a regime of large thermopower~\cite{ter97}. The nominal low-spin oxidation state of cobalt in this compound amounts to Co$^{(4-x)+}$ with occupation $3d^{5+x}$. At low energy, the electronic states are mainly governed by the $t_{2g}$ manifold of the Co$(3d)$ shell, which
becomes completely filled in the band-insulating $x$=1 limit. With a bandwidth $W$$\sim$1.6 eV~\cite{sin00} and an estimated on-site Hubbard interaction $U$ of $\sim$3$-$5eV~\cite{has04,kro06}, the material is by all means located in the strongly correlated large $U$/$W$ regime.

The appropriate theoretical tool to deal with strong correlation physics is the dynamical mean-field theory (DMFT). The formation of renormalized quasiparticles, spectral-weight transfer to Hubbard excitations, local-moment behavior and the Mott transition are well described by means of a local but frequency dependent electronic self-energy $\Sigma(\omega)$~\cite{Georges96}.
In sodium cobaltate, most of the puzzling correlation physics however appears in the strongly doped regime for $x\gtrsim0.6$~\cite{li04}. Some experiments suggest that even the superconducting dome occurs in the very same part of the phase diagram because of Co-charging effects via H$_2$O~\cite{oht11}. Here the in-plane magnetic characteristics change from antiferromagnetic (AFM) tendencies towards FM~\cite{lan08,lec09}. We will see that, despite the high doping level, nonlocal corrections to DMFT become important. Incorporating the impact of the long-range FM fluctuations around the $\Gamma$-point
is naturally hard to achieve in cluster extensions to DMFT. Not so for novel diagrammatic extensions thereof, such as the dual (fermion/boson)~\cite{rub08,rub12} approaches or D$\Gamma$A~\cite{tos07}.
In this work, we promote these techniques to a new level by combining the dual-fermion method with an ab-initio approach.
This way, we are able to show that the interaction of quasiparticles with collective magnetic excitations can lead to intriguing many-particle excitations in real materials prone to magnetic order.
As our main result, we identify previously observed excitations in correlated Na$_x$CoO$_2$ close to in-plane FM order as intricate spin-polaron excitations.

Several works have shown~\cite{mot04,hae06,lec09,boe12} that, due to the $t_{2g}$ filling and trigonal crystal-field effects, the bulk of the essential cobaltate physics may be described already within an interacting effective one-band model of dominant $a_{1g}$ character.
Figure~\ref{material-fig}b displays the band structure in the local density approximation to density functional theory for sodium cobaltate at $x$=0.67. Besides the isolated $t_{2g}$-like manifold at low energy we readily see the Fermi-surface-forming maximally-localized Wannier-like band with dominant $a_{1g}$-like character. Note that the $e_g'$-like hole pockets are filled at this doping. To good approximation, the corresponding bands may be ignored for the specific low-energy physics, leading to an effective one-band tight-binding description. For the hopping amplitudes $(t, t', t'')$ up to third nearest neighbor (NN) distances we obtain values of (-178, 41, 27) meV at  $x$=0.67.
The real-space Wannier orbital with local $a_{1g}$ resemblance in Fig.~\ref{material-fig}c has not only significant weight on the nearby oxygen ions. It also displays a mixed $e_g/e_g'$ structure on NN Co ions, rendering the entangled multi-orbital contributions obvious. Including electron hopping up to at least the third-nearest neighbors in subsequent many-body treatments of Na$_x$CoO$_2$ is crucial. The long-range hopping is essential to describe the band flattening close to the $\Gamma$-point, resulting in an extended Van-Hove singularity at the upper band edge for the cobaltate ($t$$<$0) case. Since we do not expect any qualitative changes, the full Wannier Hamiltonian for $x$=0.67 is used for the kinetic part in the many-body calculations at all investigated doping levels.

With the derived Wannier dispersion $\varepsilon(\kv)$, the complete interacting problem is cast into an effective one-band Hubbard model on the triangular CoO$_2$ lattice, i.e.
\begin{equation}
H=\sum\limits_{\kv\sigma}\varepsilon(\kv)
c_{\kv\sigma}^\dagger c_{\kv\sigma}+\sum\limits_{i}Un_{i\uparrow}n_{i\downarrow}\;.
\end{equation}
Here $\kv$ denote quasi-momenta and the index $i$ labels the lattice sites. The operators $c^{(\dagger)}_{\sigma}$ denote annihilators (creators) for the Wannier electrons with spin projection $\sigma=\uparrow,\downarrow$ and we write $n_{\sigma}$=$c_{\sigma}^{\dagger}c_{\sigma}$.
In line with previous experimental and theoretical studies~\cite{has04,kro06}, we choose an on-site Coulomb interaction of $U$=5 eV.
The data discussed in the following is obtained at temperature $T$=387K.
At various doping levels $x$, we apply the dual fermion (DF) approach~\cite{rub08} using ladder summation~\cite{haf09} to solve this realistic many-body problem tailored to the key Na$_x$CoO$_2$ physics.

In the dual fermion (DF) approach, the collective two-particle excitations are described by the Bethe-Salpeter equation (BSE) for the lattice vertex function $\Gamma$, which reads
\begin{equation}
[\tilde{\Gamma}_{\nu}^{\text{s/c}}]^{-1}_{\omega\omega'}(\qv)=
[\gamma_{\nu\,}^{\text{s/c}}]^{-1}_{\omega\omega'} -
\tilde{\chi}_{\nu}^{\omega}(\qv)\delta_{\omega\omega'}\;.
\end{equation}
Here $\omega_{n}$=$(2n+1)\pi/\beta$ and $\nu_{m}$=$2m\pi/\beta$ denote the fermionic and bosonic Matsubara frequencies and the equation is valid for the spin (s) and charge (c) channels. We use the tilde to indicate that the quantities are defined in terms of dual fermions.
Here $\gamma$ plays the role of the irreducible vertex and is given by the fully connected vertex function of the DMFT impurity model. It is local, but frequency dependent. The frequency dependence is important for the description of the collective excitations when correlations are strong. The BSE generates the sum of ladder diagrams at all orders and describes repeated particle-hole scattering which gives rise to long-wavelength two-particle collective excitations~\cite{Hafermann14-2}.

In DF, we can describe the scattering of single-particle with the collective excitations through a momentum dependent (dual) electronic self-energy $\tilde{\Sigma}_{\omega}(\kv)$.
Using the shorthand notation $\hat{\Gamma}^{\text{c/s}}_{\nu \omega \omega'}(\qv)$=$\tilde{\Gamma}^{\text{c/s}}_{\nu \omega \omega'}(\qv)-\frac{1}{2}\gamma^{\text{c/s}}_{\nu \omega \omega'}$, it can be written as
\begin{equation}
\label{sigma}
\tilde{\Sigma}_{\omega}(\kv) = \sum_{\alpha\nu\omega'\qv} a_{\alpha}
\gamma^{\nu\,\alpha}_{\omega\omega'}
\tilde{G}_{\omega+\nu}(\kv+\qv)\tilde{\chi}_{\nu}^{\omega'}(\qv)
\hat{\Gamma}^{\nu\,\alpha}_{\omega'\omega}(\qv)\;,
\end{equation}
where the sum over $\alpha$ runs over the charge and spin channels and $a_{\text{c}}$=$1/2$,
$a_{\rm s}=3/2$. The frequency (momentum) sums are understood to be normalized by the inverse temperature $\beta=1/T$ and number of k-points, respectively. The lattice Green's function, which can be related to observables, is expressed in the following form~\cite{rub12}
\begin{equation}
G_{\omega}(\kv)=[g_{\omega}^{-1}(1+\tilde{\Sigma}_{\omega}(\kv) g_{\omega})^{-1}+\Delta_{\omega}
-\varepsilon(\kv)]^{-1}\;.
\end{equation}
\begin{figure}[t]
\begin{center}
(a)\hspace*{-0.025cm}\includegraphics*[height=6.5cm]{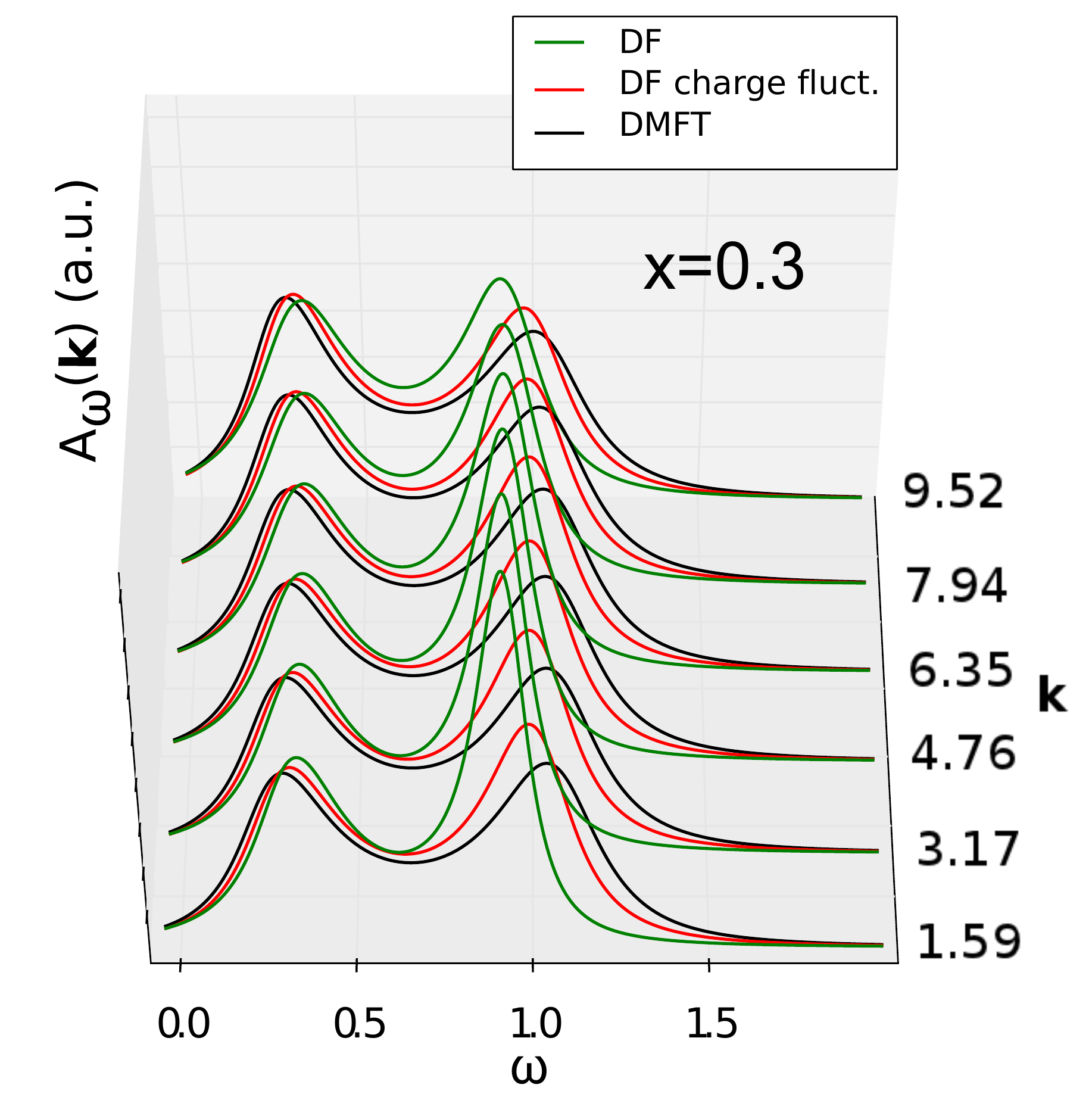}\\
(b)\hspace*{-0.025cm}\includegraphics*[height=6.5cm]{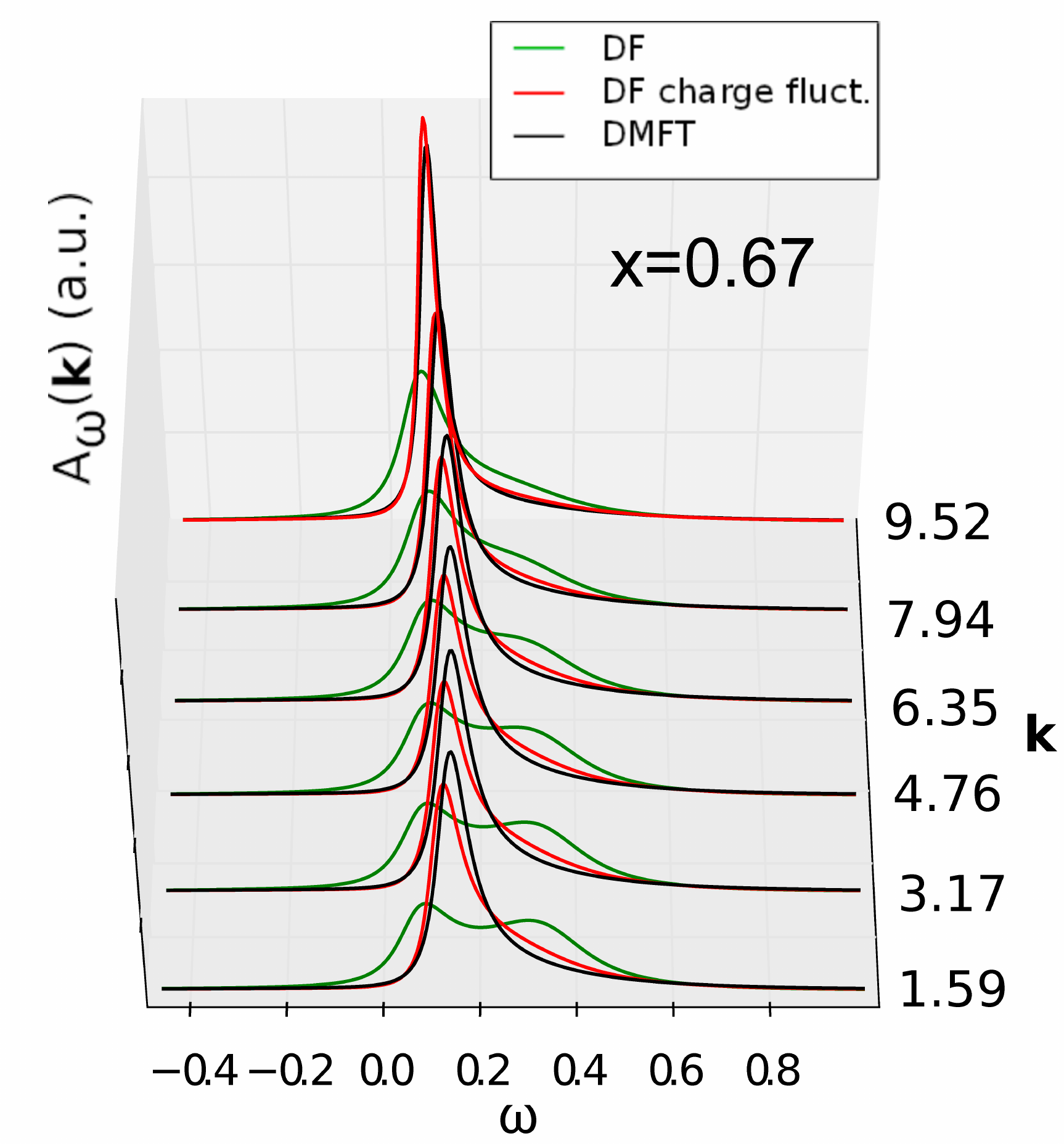}
\end{center}
\caption{(Color online) Interacting momentum-resolved Na$_x$CoO$_2$ spectral functions close to the $\Gamma$ point in the direction $\Gamma\to\text{M}$ and in units of $10^{-2}\, (2\pi/a)$. The results are for doping levels $x$=0.3 (a) $x$=0.67 (b). Results for DMFT, dual fermion (DF) and DF restricted to charge fluctuations are shown. The Fermi level is located at $\omega$=0.
\label{dual-03067-fig}
}
\end{figure}
In the above, $\tilde{\chi}_{\nu}^{\omega}(\qv)$=$-\sum_{\kv}\tilde{G}_{\omega}(\kv)\tilde{G}_{\omega+\nu}(\kv+\qv)$ denotes the dual particle-hole bubble, where $\tilde{G}_{\omega}(\kv)=G^{\text{DMFT}}_{\omega}(\kv)-g_{\omega}$. Subtracting the local self-consistent impurity Green's function $g_{\omega}$ from the DMFT lattice Green's function $G^{\text{DMFT}}_{\omega}(\kv)=[g_{\omega}^{-1}+\Delta_{\omega}-\varepsilon_{\kv}]^{-1}$ efficiently avoids double counting of local contributions.
The local one- and two-particle Green's functions to be determined within the DF method as well as in comparative single-site DMFT calculations are computed within a continuous-time quantum Monte Carlo scheme~\cite{Gull11} (see supplement).
It is known that charge-ordering effects may influence the triangular hopping paths~\cite{pei11}. However
nonlocal charge correlations are included in the DF approach and two-particle spin and
charge susceptibilities are indeed very similar to recent DMFT+vertex results~\cite{boe12}.
The principal physics discussed here is not altered by the presence of charge ordering.

Figure~\ref{dual-03067-fig} shows the DF and DMFT spectral functions $A_{\omega}(\kv)=-(1/\pi)\Im G_{\omega+i0^{+}}(\kv)$ in the unoccupied part of the spectrum $\omega>0$ in reciprocal space and close to the $\Gamma$ point. At low doping $x=0.3$, relatively close to half filling, both DMFT and DF agree qualitatively rather well, displaying a broadened quasiparticle (QP) peak and an upper
Hubbard band located at an energy of $\omega\sim 1$ eV. Within DF, the upper Hubbard band is shifted to slightly lower energies and considerably narrowed. Because of the diagrammatic construction of the the dual self-energy \eqref{sigma}, we can readily separate contributions from collective spin and charge excitations. Restricting the DF calculation to charge fluctuations results in a spectrum that is much closer to DMFT. At this doping, the magnetic susceptibility is peaked at the K-point of the Brillouin zone~\cite{boe12}, which we confirm in our data. There is hence a tendency to antiferromagnetic ordering, but the system does not order because of frustration. Instead, the presence of dynamical antiferromagnetic correlations decreases the fluctuations and leads to an interplay of Slater and Mott physics that increases the coherence of the single-particle excitations. A similar situation occurs in the two-dimensional square lattice~\cite{Hafermann12-2}.

At $x$=0.67, the difference between DMFT and DF is much more striking, giving rise to a qualitatively different excitation spectrum above $\varepsilon_{\rm F}$. In DMFT, the quasiparticle peak is considerably larger and the QP weight $Z$ is significantly enhanced compared to the case of low doping. This is expected, since $Z\approx 1$ should hold far away from half filling and towards the opposite (band-insulating) endpoint $x$=1. We further see that the upper Hubbard band has completely dissolved in the DMFT perspective. In DF, on the contrary, the QP peak close to $\Gamma$ is strongly renormalized. The spectral function additionally exhibits a broad sideband excitation at $\omega\sim0.3-0.4$~eV. By restricting the DF calculation to the charge channel only, this sideband excitation disappears. This is a strong indication that this excitation is of \emph{magnetic} origin.

A weak absorption feature already present at room temperature has previously been reported from optics experiments of nearly-ferromagnetic Na$_{0.7}$CoO$_2$ for $x$=0.7 by Wan {\sl et al.}~\cite{wan04}. The broad feature was observed in the mid infrared at an energy of $\omega$$\sim$$0.4$ eV, which is in remarkably good agreement with our result. The authors of that work speculated about spin-polarons as one of the possible mechanisms.
This interpretation is not unlikely due to an enhanced ferromagnetic susceptibility and the presence of strong FM fluctuations at this doping. We find that
the leading eigenvalue of the magnetic channel of the Bethe-Salpeter equations is largest for $\qv=0$, i.e. for ferromagnetic alignment of the spins. This is in line with paramagnons found in the vicinity of the $\Gamma$ point~\cite{boe12}.

In elementary theory for the spin-polaron, one considers an electron in a ferromagnetic background and a single magnon in the $s$$-$$d$~\cite{Methfessel68,Shastry81,Auslender84} or $t$$-$$J$ models~\cite{Katsnelson82}.
In the latter, the presence of bound states depends crucially on the lattice dimensionality and anomalies in the electronic spectrum. In three dimensions (3D) and small FM $J>0$, there are no spin-polaron bound states in an almost filled band in presence of Nagaoka ferromagnetism~\cite{[{The Nagaoka ferromagnet is a saturated ferromagnetic state within the nearly half-filled infinite-$U$ Hubbard model. See }][{}]nag66,*irk85}. This is similar to the classical Slater-Koster impurity problem, where a weak potential in a 3D crystal does not lead to the formation of a bound state. In 1D however, a bound state forms at arbitrarily small ratio of $J/t$~\cite{Katsnelson82}.
For the quasi-2D system the situation is marginal and depends on the existence of the Van-Hove singularity in the band structure. In order to obtain a qualitative description of spin-polarons on the present triangular lattice, we consider a numerical solution of the $t-J$ model. 
The problem becomes tractable by restricting it to a state of a single magnon and an excess charge carrier in the ferromagnetic state denoted $|{\rm FM}\rangle$: $c_{i\sigma}^{\dagger}S_{j}^{-}|{\rm FM}\rangle$ ($S_{j}^{-}$ lowers the spin at site $j$ by 1). This allows us to compute the dispersion $E(\kv)$ of the resulting bound state numerically (see the supplementary material for details).

\begin{figure}[t]
\begin{center}
\includegraphics*[width=0.47\textwidth]{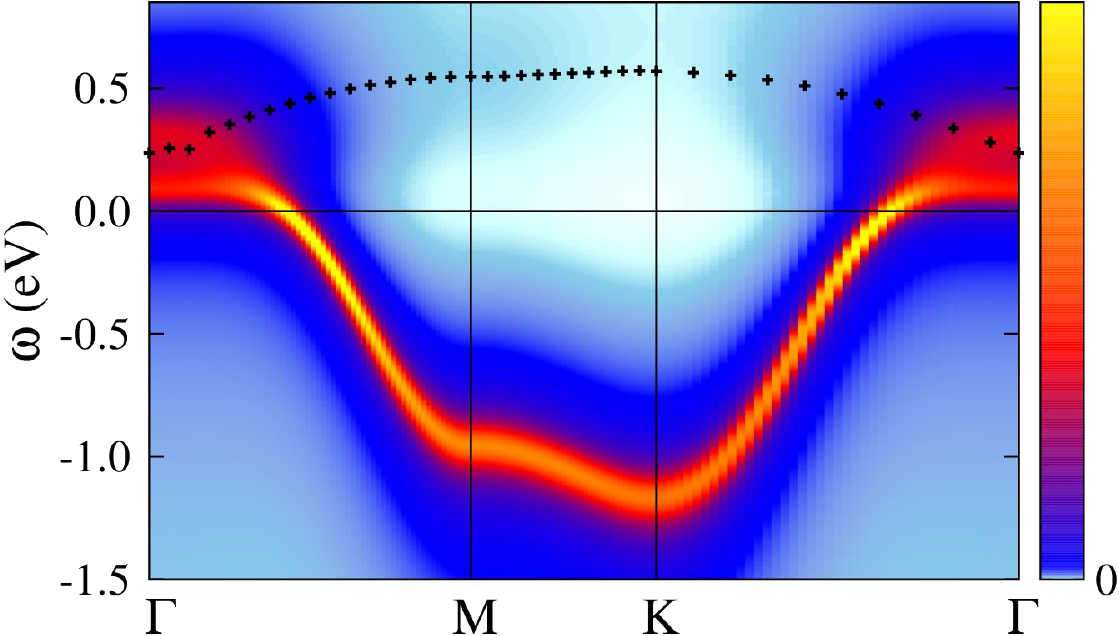}
\end{center}
\caption{(Color online) Intensity plot of the interacting $k$-resolved Na$_x$CoO$_2$
spectral function for $A_{\omega}({\bf k})$ (in arbitrary units) for doping x=0.67 along the high-symmetry lines through $\Gamma=(0,0)$, M$=(1/2,-1/2\sqrt{3})$ and K$=(2/3,0)$ (in units of $2\pi/a$). The spin-polaron states close to $\Gamma$ are clearly visible. Dark crosses show the solution for the spin-polaron band in a $t$$-$$J$ model.
\label{dfspecfig}}
\end{figure}

In Fig.~\ref{dfspecfig} we display the momentum-resolved DF spectral function for $x=0.67$ together with the model spin-polaron dispersion $E({\bf k})$.
We see an overall moderate bandwidth renormalization in DF due to the high doping level. Above the scattering states near $\Gamma$, we recognize the sideband of bound states. The qualitative description of the spin-polaron excitations through the simplified model is remarkable. The spectral weight of the realistic calculation diminishes, but clearly extends in the same direction as the model dispersion as one moves away from $\Gamma$. Given the rudimentary nature of the model we did not attempt to fit the value of the effective exchange $J$. Nevertheless, the value of $J=0.1\abs{t}=17.8$meV we used here has the same order of magnitude and --given the simplicity of the model-- is quite close to experimental results. Spin-model analyses of inelastic neutron data for Na$_{x}$CoO$_2$ at high $x$ yield $J$$\sim$5-6 meV for the intra-layer ferromagnetic spin coupling~\cite{bay05,hel06}.
Note that the effective $J$ is not given by the (AFM) super-exchange $-4t^{2}/U$ valid in the spin-1/2 Heisenberg-limit of the Hubbard model close to half filling.

In order to better understand these results, it is instructive to scrutinize the mechanism by which the spin-polarons emerge in the DF calculation.
We recall that the self-energy, Eq.~\eqref{sigma}, has the form of a convolution: $\tilde{\Sigma}_{\omega}(\kv)=\sum_{\nu\qv}\tilde{G}_{\omega+\nu}(\kv+\qv)S_{\nu}(\qv)$, where $S_{\nu}(\qv)$ encodes information on the magnetic excitations at momentum $\qv$. The dominant contribution to $\Sigma$ near $\Gamma$, which is responsible for the bound states, stems from those terms in the sum over $\qv$ for which the product of $\tilde{G}(\kv+\qv)$ and $S(\qv)$ dominates. Due to the presence of the Van-Hove singularity, the density of states is large for $\kv+\qv\approx\mathbf{0}$, or $\abs{\kv}\approx\abs{\qv}\approx\mathbf{0}$. As mentioned previously, the susceptibility also dominates for $\qv\approx\mathbf{0}$. The self-energy is thus dominated by the ferromagnetic spin excitations at this wave vector.

The bound states in the many-body calculation hence emerge from ferromagnetic paramagnons. This explains the two most striking aspects of our results: In agreement with the experiment, the sideband is rather broad~\cite{wan04}. The quick decay of these excitations is a result of the fact that the lifetime of both the magnon excitations and the quasiparticles is finite. In the simple model, the excitations have infinite lifetime due to the absence of decay channels.
Secondly, their intensity vanishes quickly when moving away from $\Gamma$. According to the foregoing, this is a direct consequence of the fact that at $x$=0.67 and elevated temperature, well-defined ferromagnetic paramagnons only exist in close vicinity of the $\Gamma$-point~\cite{boe12}.

We hence identify the sideband excitation as a spin-polaron excitation originating from scattering between quasi-particles and ferromagnetic magnons. The presence of the Van-Hove singularity at the top of the hole-like band is crucial, because it effectively reduces the dimensionality of the problem. It provides a large density of charge carriers and brings the system close to a Stoner instability, which therefore exhibits a tendency towards ferromagnetism.
Note that for cobaltate, we have $t$$<$0. For the opposite sign, the Van-Hove singularity occurs at the bottom of the band.  Thus the formation of a spin-polaron state of the type described in Ref.~\cite{Katsnelson82} is once more a consequence of the particular electronic structure of sodium cobaltate~\cite{lec09,pei11,boe12}.

To summarize, we have investigated the spectral properties of the doped Mott material Na$_x$CoO$_2$ within a combined first-principles many-body description. The low-doping regime is rather well described within DMFT, although dynamical antiferromagnetic correlations lead to an interplay of Slater and Hubbard physics and increased coherence of the high-energy excitations.
For higher doping we found sideband excitations at $\omega\sim 0.35$ eV in close agreement with optics experiments.
We have identified these as spin-polaron excitations resulting from bound states of quasiparticles with ferromagnetic paramagnons. To our knowledge, this is the first theoretical description of spin-polaron states emerging from \emph{dynamical ferromagnetism}.
The presence of the extended Van-Hove singularity in the realistic tight-binding spectrum is crucial for this effect.
We note that the impact of magnetic fluctuations onto charge transport in sodium cobaltate has already been described in a simplified model picture in Ref.~\onlinecite{liu04}. 
The spin-polaron physics revealed by us is qualitatively different from spin(-orbital) polarons as used by Khaliullin {\sl et al.}~\cite{kha08} to describe strongly renormalized bands in the  occupied part of the spectrum of strongly doped sodium cobaltate. There, orbital-dependent scattering involving the onsite Hund's $J_H$ is a vital ingredient. Here the much smaller inter-site $J$ is the driving force for the formation of the spin-polarons.
It is worthwhile to further study these many-body quasi-bound states theoretically and experimentally. The latter may be accomplished by high-resolution inverse photoemission experiments or quasiparticle-scattering analysis within scanning-tunneling-microscopy.

\begin{acknowledgments}
We would like to thank A. Rubtsov, S. Brener and L. Boehnke for helpful discussions.
This work has been supported by the DFG-FOR1346 program. M. I. K. acknowledges financial support from ERC (project 338957 FEMTO/NANO) and from NWO via the Spinoza Prize and H. H. from the FP7/ERC, under Grant Agreement No. 278472-MottMetals. Computations have been performed using high-performance computing resources at the NIC, Forschungszentrum J\"ulich, under project HHH14. The impurity solver~\cite{Hafermann13} and the dual fermion implementation are based on the ALPS libraries~\cite{ALPS2}.

\end{acknowledgments}

\bibliographystyle{apsrev4-1}
\bibliography{bibextra}

\clearpage

\begin{center}
{\LARGE Supplemental Information}
\end{center}

\section{Band structure calculations}

Density functional theory in the local density approximation (LDA) is employed for the material-specific part. To achieve finite doping in the pseudopotential approach, an effective Na~ion with nuclear charge $Z_{\rm Na}$$+$$x$ is used in virtual-crystal LDA calculations for a single-formula-unit cell.
Interlayer bonding-antibonding splittings are thereby neglected. For the lattice parameters
at $x$=0.67 the values $a$=2.83\AA$\,$ and $c$=10.90\AA$\,$ (given in double-layer
counting) are used. We utilize the relaxed LDA value $z_0$=0.084 for the O height as in the
work by Singh~\cite{sin00}.

\section{Dual fermion calculations}

The dual fermion approach~\cite{rub08} is a diagrammatic extension of dynamical mean-field theory. As in DMFT, the approach is based on a quantum impurity model subject to a self-consistency condition. In addition to the impurity Green's function and self-energy, the impurity vertex function has to be computed.
We employ a hybridization expansion continuous-time quantum Monte Carlo solver~\cite{Hafermann13} with improved estimators~\cite{Hafermann12} for the self-energy and vertex for this purpose.
These quantities enter the dual Green's function and interaction. The non-local dual self-energy is evaluated within a Feynman-type diagrammatic expansion and subsequently transformed to obtain the physical self-energy or Green's function. Here we used the ladder approximation~\cite{haf09} in order to capture the effect of the long-range collective excitations.

For the Monte Carlo as well as for the dual fermion part for retrieving the interacting Green's functions, we employ a fully parallelized implementation. We exploit the lattice symmetries and compute the diagrams on a $128\times 128\times128$ frequency-lattice grid using the fast-Fourier-transform algorithm. For further details on the dual fermion formalism and dual perturbation theory, see Ref.~\cite{Hafermann12-2}.

\section{Bound states of electron and magnon on arbitrary lattices}

We consider a ferromagnetic $t$$-$$J$ model given by the Hamiltonian
\begin{equation}
  H=\sum\limits_{ij\sigma} t_{ij}
c_{i\sigma}^{\dagger}c_{j\sigma} -\frac{1}{2}\sum\limits_{ij}
J_{ij} \vec{S}_{i}\vec{S}_{j}\;,
\end{equation}
where $c_{i\sigma}^{(\dagger)}$ destroys (creates) a particle with spin-projection $\sigma$
on the site $i$ only if there is already a particle with opposite spin. $J_{ij}>0$ corresponds to ferromagnetic coupling.
We call the state with
every site occupied by one electron with spin-up $|{\rm FM}\rangle$. In the following, we restrict
the investigation to the specific sector $c_{i\sigma}^{\dagger}S_{j}^{-}|{\rm FM}\rangle \equiv |ij\rangle$.

With the ansatz
\begin{equation}
  |\Psi\rangle=\sum\limits_{ij,i\not=j}\psi_{ij}|ij\rangle
\end{equation}
and the eigenvalue equation $H|\Psi\rangle=E|\Psi\rangle$ it follows for the coefficients
$\psi_{ij}$~\cite{Katsnelson82}
\begin{equation}
  E \psi_{ij} = \sum\limits_{k\not=j}t_{ik}\psi_{kj} +
t_{ij}\psi_{ji} + \sum\limits_{k\not=i}J_{jk}(\psi_{ij}-\psi_{ik}). \label{eq:schroedinger}
\end{equation}
From the definition of $|ij\rangle$ it is clear that $|ii\rangle$=0 holds. We therefore
define $\psi_{ii}$=0. With the Fourier expansion of the form
\begin{align}
  \psi_{ij}&=\sum\limits_{\vec{q}}\psi_{\vec{q}} \exp{\left[i\left(\vec{q}\vec{R}_i +
\left(\vec{Q}-\vec{q}\right)\vec{R}_j\right)\right]}, \label{eq:fourrierexpansion1}\\
  \psi_{\vec{q}}&=\sum\limits_{ij}\psi_{ij} \exp{\left[-i\left(\vec{q}\vec{R}_i +
\left(\vec{Q}-\vec{q}\right)\vec{R}_j\right)\right]}, \label{eq:fourrierexpansion2}
\end{align}
we restrict the solutions to the subspace of electron-magnon pairs propagating with the
wave vector $\vec{Q}$. Transforming (\ref{eq:schroedinger}) via (\ref{eq:fourrierexpansion2})
leads to~\cite{Katsnelson82}
\begin{eqnarray}
  \left(E - J_0 - t_{\vec{q}} + J_{\vec{Q} - \vec{q}} \right) \psi_{\vec{q}}&=&\nonumber\\
&&\hspace*{-3cm} \sum_{\vec{p}} \left( t_{\vec{Q}-\vec{q}-\vec{p}} - t_{\vec{p}} +
J_{\vec{Q}-\vec{p}} - J_{\vec{q}-\vec{p}} \right) \psi_{\vec{p}}\;, \label{eq:fredholminteq}
\end{eqnarray}
with $J_0$=$J_{\vec{q}=\vec{0}}$. This is a Fredholm integral equation of the second kind with
separable integral kernel (if we take into account the hopping and exchange to a finite number of neighbors in real space).
To solve that integral equation we separate the kernel
\begin{equation}
  K_{\vec{q}\vec{p}}=t_{\vec{Q}-\vec{q}-\vec{p}} - t_{\vec{p}} + J_{\vec{Q}-\vec{p}} -
J_{\vec{q}-\vec{p}}
\end{equation}
by rewriting it as
\begin{align}
  K_{\vec{q}\vec{p}}&=\sum\limits_{\vec{R}} \left[ t_{\vec{R}}\left
(e^{i \left( \vec{Q}-\vec{q}-\vec{p} \right) \vec{R}} - e^{i \vec{p} \vec{R}} \right)\right.\notag\\
 &\qquad +\left. J_{\vec{R}}\left(e^{i \left( \vec{Q}-\vec{p} \right) \vec{R}} - e^{i \left( \vec{q} -
\vec{p} \right) \vec{R}} \right) \right] \\
  &=\sum\limits_{\vec{R}} \left[ t_{\vec{R}}\left(e^{i \left( - \vec{Q} + \vec{q} \right) \vec{R}} - 1 \right) \right. \notag\\
&\qquad +\left.
J_{\vec{R}}\left(e^{-i \vec{Q} \vec{R}} - e^{-i \vec{q} \vec{R}} \right) \right]e^{i\vec{p}\vec{R}} \notag \\
  &=\sum\limits_{\vec{R}} N_{\vec{R}}(\vec{p})M_{\vec{R}}(\vec{q})\;,
\end{align}
where
\begin{align}
  N_{\vec{R}}(\vec{p})&=e^{i\vec{p}\vec{R}} \\
  M_{\vec{R}}(\vec{q})&=t_{\vec{R}}\left
(e^{i \left( - \vec{Q} + \vec{q} \right) \vec{R}} - 1 \right) +
J_{\vec{R}}\left(e^{-i \vec{Q} \vec{R}} - e^{-i \vec{q} \vec{R}} \right)\;.
\end{align}
After dividing (\ref{eq:fredholminteq}) by the factor on the lefthand side, i.e.
\begin{equation}
  \psi_{\vec{q}} = \sum\limits_{\vec{R}} \frac{M_{\vec{R}}(\vec{q})}
{E - J_0 - t_{\vec{q}} + J_{\vec{Q} - \vec{q}}}
\sum\limits_{\vec{p}}N_{\vec{R}}(\vec{p}) \psi_{\vec{p}}\;,
\end{equation}
we multiply both sides with $N_{\vec{R'}}(\vec{q})$ and sum over $\vec{q}$:
\begin{equation}
  \sum\limits_{\vec{q}}N_{\vec{R}'}(\vec{q})\psi_{\vec{q}} =
\sum\limits_{\vec{R}}\sum\limits_{\vec{q}} \frac{N_{\vec{R}'}
(\vec{q})M_{\vec{R}}(\vec{q})}{E - J_0 - t_{\vec{q}} + J_{\vec{Q} - \vec{q}}}
\sum\limits_{\vec{p}}N_{\vec{R}}(\vec{p})\psi_{\vec{p}}\;.
\end{equation}
This can be written in the form
\begin{equation}
  c_{\vec{R}'} = \sum\limits_{\vec{R}} a_{\vec{R}'\vec{R}} c_{\vec{R}}\; \label{eq:eigenvalue0}
\end{equation}
with
\begin{align}
  c_{\vec{R}}&=\sum\limits_{\vec{q}}N_{\vec{R}}(\vec{q})\psi_{\vec{q}} \label{eq:cr},\\
  a_{\vec{R}'\vec{R}}&=\sum\limits_{\vec{q}}
\frac{N_{\vec{R}'}(\vec{q})M_{\vec{R}}(\vec{q})}
{E - J_0 - t_{\vec{q}} + J_{\vec{Q} - \vec{q}}}\;. \label{eq:amatrix}
\end{align}
Or in matrix form,
\begin{equation}
\label{matrix}
  \left( \mathbbm{1} - A(E) \right) \vec{C}=0\;,
\end{equation}
where $A_{ij}=a_{\vec{R}_{i}\vec{R}_{j}}$ and $C_{i}=c_{\vec{R}_{i}}$.
The dispersion $E_{\vec{Q}}$ of the spin-polaron band for given $\vec{Q}$ is determined by solving $\text{det}(\mathbbm{1} - A(E))$=0 numerically for the energy $E_{\vec{Q}}$ and with the constraint $|E_{\vec{Q}} - J_0|$$>$$-t_{\vec{q}} + J_{\vec{Q} - \vec{q}}$.

In one dimension, an exact solution exists~\cite{Katsnelson82}:
\begin{align}
E^{\rm (1D)}(k)=-\frac{2t[t-J\cos(k)]}{|t|+J}\;.
\end{align}
For the two-dimensional case, we have solved Eq.~\eqref{matrix} on the triangular lattice. We replace $t_{ij}$ by the values $t$, $t'$ and $t''$ given in the main text as appropriate. We further use an effective exchange constant $J_{ij}=J>0$ for $i$ and $j$ nearest neighbors and $J_{ij}=0$ otherwise. The resulting matrix has size $18\times 18$. The number $18$ is due to the fact that the sum in \eqref{eq:eigenvalue0} is over the three shells of nearest-, next-nearest and third-nearest neighbors, which on the triangular lattice contain 6 atoms each.

\end{document}